\documentclass[runningheads,fleqn]{svmult}

\usepackage{makeidx}   
\usepackage{graphicx}  
\usepackage{multicol}  
\usepackage{physproc}  
\makeindex             

\begin{document}
\title*{Chiral Symmetry Versus the Lattice}
\toctitle{Chiral Symmetry Versus the Lattice}
%
%
\titlerunning{Chiral Symmetry Versus the Lattice}
%
\author{Michael Creutz}
\authorrunning{Michael Creutz}
%
%
\institute{Physics Department, Brookhaven National Laboratory, Upton,
NY 11973, USA}
\maketitle              

\begin{abstract}
After mentioning some of the difficulties arising in lattice gauge
theory from chiral symmetry, I discuss one of the recent attempts to
resolve these issues using fermionic surface states in an extra
space-time dimension.  This picture can be understood in terms of end
states on a simple ladder molecule.
\end{abstract}

Chiral symmetry and lattice gauge theory provide two of the best known
approaches to understanding non-perturbative phenomena in relativistic
quantum field theory.  However, rather interesting clashes appear when
these methods are considered together.  Our understanding of this
problem has seen considerable progress in the last few years, although
numerous unanswered questions remain.  The purpose of this talk is to
introduce some of these issues from a general point of view, avoiding
technical details.  For a more extensive reviews, see
\cite{Creutz:2000bs} and \cite{neuberger}.

The issues involved are rather old, going back to the species doubling
phenomena observed with the first papers on lattice gauge theory.  To
solve this doubling, the regulator was modified, but this modification
directly breaks chiral symmetry.  This feature is not a nemesis, but a
virtue of the formalism.  Without such modifications, there would be
no room for the well known chiral anomalies to appear.  Indeed, I
believe that the conflict between chiral symmetry and the lattice is
telling us something deep about the structure of relativistic quantum
field theory.

\begin{figure}[b]
\centerline{\includegraphics[width=.3\textwidth]{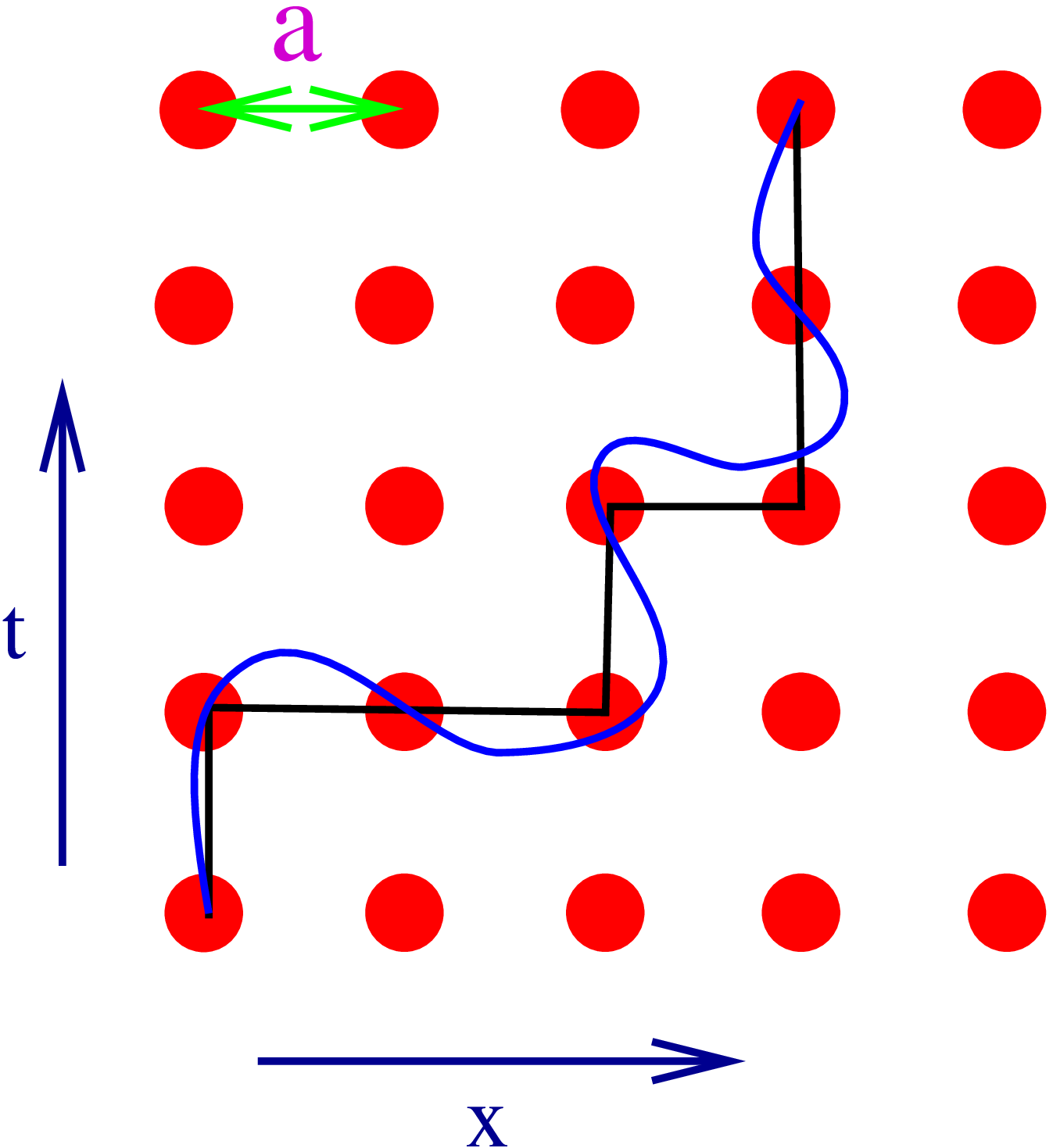}}
\caption[]{Lattice gauge theory begins by approximating continuous
space time with a discrete set of points}
\label{worldline}
\end{figure}

I begin with a brief reminder of what lattice gauge theory is all
about.  Basically, it is nothing but a mathematical trick.  By
removing the infinities of the underlying field theory, the lattice
gives us a well defined mathematical system independent of
perturbative expansion.  In this approach the world lines of a
particle are replaced by discrete hops on a four dimensional lattice,
as sketched in Fig.~\ref{worldline}.

The lattice spacing {$a$} is an artificial construct and we must
always keep in mind the need to take {$a\rightarrow 0$} for physical
results.  While in place, however, the lattice provides an ultraviolet
cutoff at momentum {$\Lambda = \pi/a$}.  In addition to making the
theory finite, the lattice enables Monte Carlo simulations, which
currently dominate the field.

Since the lattice is a first principles approach to field theory, one
could ask why care about the details of chiral symmetry.  Just put the
problem on the computer, predict particle properties, and they should
come out correctly if the underlying dynamics is relevant.  While this
may perhaps be a logical point of view, it ignores a vast lore built
up over the years.  In the context the strong interactions, the pion
and the rho mesons are made of the same quarks, the only difference
being whether the spins are anti-parallel or parallel.  Yet the pion,
at 140 MeV, weighs substantially less than the 770 MeV rho.  Chiral
symmetry is at the core of the conventional explanation.  Since the up
and down quarks are fairly light, we have an approximately conserved
axial vector current, and the pion is believed to be the remnant
Goldstone boson of a spontaneous breaking of this chiral symmetry.

Another motivation for studying chiral issues arises when considering
the weak interactions.  Here we are immediately faced with the
experimental observation of parity violation, neutrinos are left
handed.  In the standard electroweak model fundamental gauge fields
are coupled directly to chiral currents.  The corresponding symmetries
are gauged, {\it i.e.} they become local, and are crucial to the basic
structure of the theory.  Since the lattice is the one truly
non-perturbative regulator for defining a field theory, if one cannot
find a lattice regularization for the standard model, the standard
model itself may not be well defined.

A third reason to explore chiral symmetries comes from unified field
theories.  These usually have a large natural scale.  In comparison,
quark and lepton masses are much smaller.  In such models chiral
symmetry can protect fermion masses from large renormalization.  This
is also one of the prime reasons for the popularity of super-symmetry,
which extends this protection to scalar particles, such as the Higgs
meson.

The word ``chiral,'' based on the Greek word for hand, was introduced
into modern scientific jargon by Lord Kelvin \cite{kelvin} in 1904
when in a rewriting of his Baltimore lectures he said ``I call any
geometrical figure, or group of points, chiral, and say it has
chirality, if its image in a plane mirror, ideally realized, cannot be
brought to coincide with itself.''

The concept of chirality is most frequently used by chemists.
Molecules whose structure is different from their mirror image are
called chiral.  For the particle theorist, however, the use of this
term is associated with subtleties of the Lorentz group and massless
particles.  When a particle is massless it travels at the speed of
light.  This is a limiting velocity for any observer, who cannot go
faster than such a particle to reverse its direction.  A direct
consequence for particles with spin is that their helicity,
i.e. angular momentum along their direction of motion, is frame
invariant.  For spin 1/2 fermions, the left and right handed
components, $\psi_L$ and $\psi_R$ become independent fields.  This
independence is naively preserved under gauge interactions; a
relativistic electron tends to preserve its helicity as it travels
through electromagnetic fields.

\begin{figure}
\centerline{\includegraphics[width=.3\textwidth]{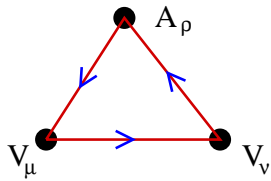}}
\caption{The triangle diagram cannot be regulated so both
vector and axial vector currents are conserved.}
\label{fig:triangle}
\end{figure}

This concept, however, is clouded by the so-called ``chiral
anomalies'' \cite{abj}.  In particular, the famous triangle diagram, sketched in
Fig.~\ref{fig:triangle}, coupling two vector and one axial vector
current is divergent, and no regularization can keep them both
conserved.  If either is coupled to a gauge field, such as
electromagnetism, this diagram must be regulated with that particular
current being conserved.  Then the other cannot be.  These anomalies
are at the core of the lattice problems.

In one spatial dimension chirality reduces to separating perticles
into left and right movers.  In this case the anomaly is easily
understood via simple band theory \cite{anomalyflow}.  A particle of
non-zero mass $m$ and momentum $p$ has energy $E=\pm\sqrt{p^2+m^2}$.
Here I use a Dirac sea description where the negative energy states
are filled in the normal vacuum.  Considering the positive and
negative energy states together, the spectrum has a gap equal to twice
the particle mass.  In the vacuum the Fermi level is at zero energy,
exactly in the center of this gap.  In conventional band theory
language, the vacuum is an insulator.

\begin{figure}
\centerline{\includegraphics[width=.6\textwidth]{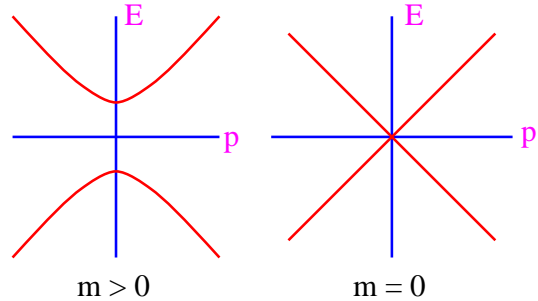}}
\caption{In one dimension the spectrum of massive particles has a gap,
and the vacuum can be regarded as an insulator.  The massless case, in
contrast, represents a conductor.  The anomaly manifests itself in the
ability to induce currents in a wire.}
\label{fig:conductor}
\end{figure}

In contrast, for massless particles where $E=\pm|p|$, the gap
vanishes.  The system becomes a conductor, as sketched in
Fig.~\ref{fig:conductor}.  Of course, conductors can carry currents,
and here the current is proportional to the number of right moving
particles minus the number of left movers.  If we consider gauge
fields, they can induce currents, a process under which the number of
right or left movers cannot be separately invariant.  This is the
anomaly, without which transformers would not work.

This induction of currents is not a conversion of particles directly
from left into right movers, but rather a sliding of levels in and out
of the infinite Dirac sea.  The generalization of this discussion to
three spatial dimensions uses Landau levels in a magnetic field; the
lowest Landau level behaves exactly as the above one dimensional
case \cite{anomalyflow}.

One particularly intriguing consequence for the standard model is that
baryon number is an anomalous charge.  Indeed, 't~Hooft\cite{thooft}
pointed out a specific baryon-number-changing mechanism through
topologically non-trivial gauge configurations.  The rate is highly
suppressed due to a small tunneling factor and is far too small to
observe experimentally.  Nevertheless, the process is there in
principle, and any valid non-perturbative formulation of the standard
model must accommodate it.  If we have a fully finite and exactly
gauge invariant lattice theory, the dynamics must contain terms which
violate baryon number.  This point was emphasized some time ago by
Eichten and Preskill \cite{eichtenpreskill} and further by Banks
\cite{banks}.

Without baryon violating terms, something must fail.  In naive
approaches to lattice fermions the problem materializes via extra
particles, the so-called doublers, which cancel the anomalies.  For
the strong interactions alone, a vector-like theory, Wilson
\cite{wilson77} showed how to remove the doublers by adding a chirally
non-symmetric term.  This term formally vanishes in the continuum
limit, but serves to give the doublers masses of order the inverse
lattice spacing.  As chiral symmetry is explicitly broken, the chiral
limit of vanishing pion mass is only obtained with a fine tuning of
the quark mass.  This is no longer ``protected''; the bare and
physical quark masses no longer vanish together.  This approach works
well for the strong interactions, but explicitly breaks a chirally
coupled gauge theory.  This entails an infinite number of gauge
variant counter-terms to restore gauged chiral symmetries in the
continuum limit \cite{rome}.  It is these features that drive us to
search for a more elegant formulation.

To proceed I frame the discussion in terms of extra space-time
dimensions.  The idea of adding unobserved dimensions is an old one in
theoretical physics, going back to Kaluza and Klein
\cite{kaluzaklein}, and often is quite useful in unifying different
interactions.  Of course the extension of space-time to higher
dimensions is crucial to modern string theories.  There are probably
further unexploited analogies here, but chiral symmetry in particular
can become quite natural when formulated on higher dimensional
membranes.  Here I use only the simplest extension, involving one
extra dimension.

\begin{figure}
\centerline{\includegraphics[width=.5\textwidth]{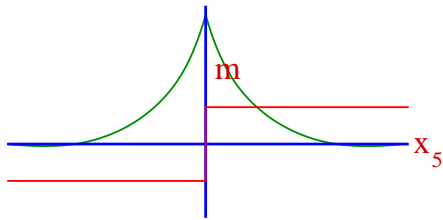}}
\caption{A step in a five dimensional fermion mass can give rise
to topological zero-energy fermion modes bound to a four dimensional 
interface.}
\label{fig:massstep}
\end{figure}

I start with an observation of Callan and Harvey \cite{callanharvey},
building on Jackiw and Rebbi \cite{jackiwrebbi}. They argue that a five
dimensional massive fermion theory formulated with an interface where
the fermion mass changes sign, as sketched in Fig.~\ref{fig:massstep},
can give rise to a four dimensional theory of massless fermionic modes
bound to the interface.  The low energy states on the interface are
naturally chiral, and anomalous currents are elegantly described in
terms of a flow into the fifth dimension.

\begin{figure}
\centerline{\includegraphics[width=.5\textwidth]{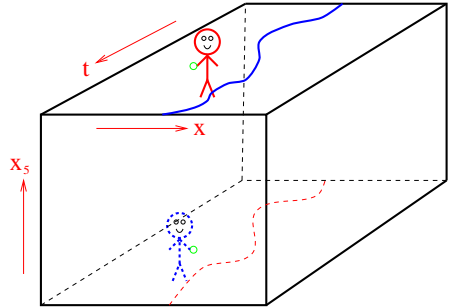}}
\caption{Regarding our four dimensional world as a surface in five
dimensions.}
\label{fig:kaplan}
\end{figure}

While the Callan and Harvey discussion is set in the continuum, Kaplan
\cite{kaplan} suggested carrying the formalism directly over to the
lattice.  In the Wilson formulation, the particle mass is controlled
via the hopping parameter, usually denoted $K$.  The massless
situation is obtained at a critical hopping, $K_c$, the numerical
value of which depends on the gauge coupling.  Thus, to set up an
interface as used by Callan and Harvey, one should consider a five
dimensional theory with a hopping parameter which depends on the extra
fifth coordinate.  This dependence should be constructed to generate a
four dimensional interface separating a region with $K>K_c$ from one
with $K<K_c$.  Shamir \cite{shamir} observed a substantial
simplification on the $K<K_c$ side by putting $K=0$.  Then that region
decouples, and the picture reduces to a four dimensional surface of a
five dimensional crystal.  The physical picture is sketched in
Fig.~\ref{fig:kaplan}.  For a Hamiltonian discussion, see
Ref.~\cite{mcih}.  Indeed, surface modes are not a particularly new
concept; in 1939 Shockley \cite{shockley} discussed their appearance
in band models when the inter-band coupling becomes strong.  This
approach has stimulated several closely related variations that have
attracted considerable recent attention \cite{Creutz:2000bs}
\cite{neuberger} \cite{nn}.

I will now discuss these ``domain-wall fermions'' from a rather
unconventional direction.  Following a recent paper of mine
\cite{icetray}, I present the subject from a ``chemists'' point of
view, in terms of a chain molecule with special electronic states
carrying energies fixed by symmetries.  For lattice gauge theory,
placing one of these molecules at each space-time site gives
excitations of naturally zero mass.  This is in direct analogy to the
role of chiral symmetry in conventional continuum descriptions.  After
presenting this picture, I will wander into some comments and
speculations about exact lattice chiral symmetries and schemes for
gauging them.

\begin{figure}
\centerline{\includegraphics[width=.7\textwidth]{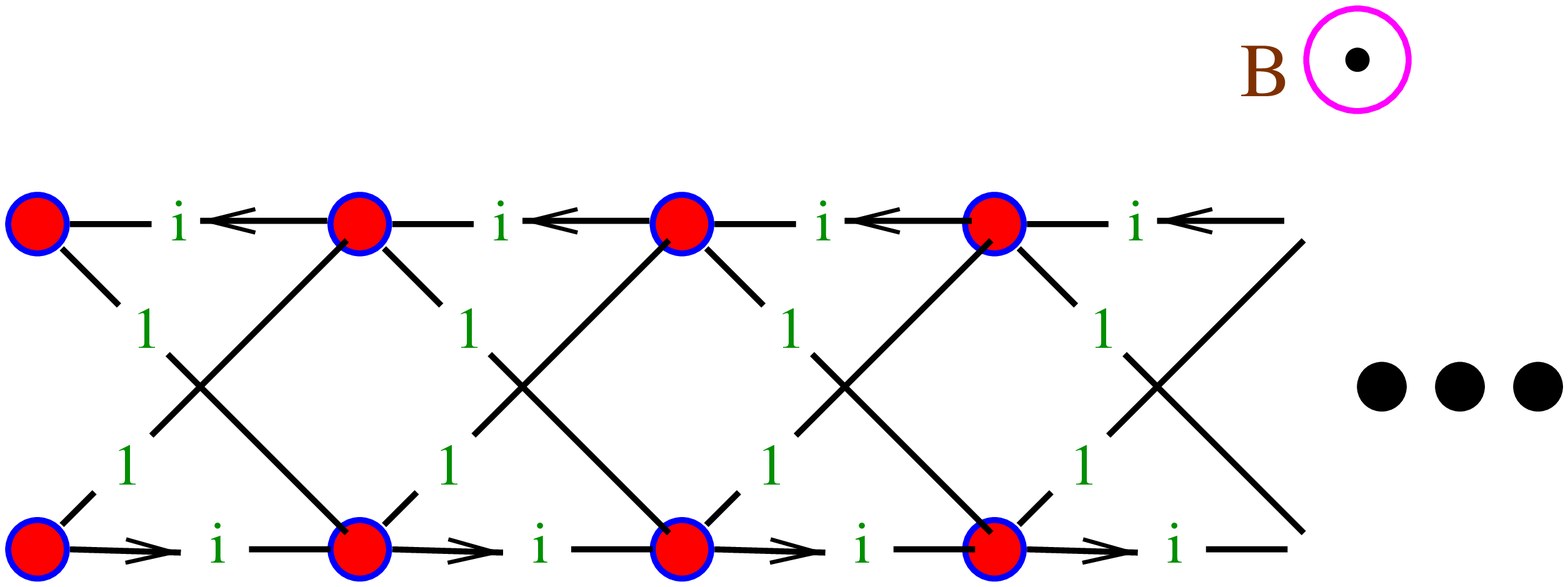}}
\caption
{The basic cross linked lattice in a magnetic field.  The numbers on
the bonds represent phases giving half a unit of flux per plaquette.
If we slightly slope the vertical bonds alternately in and out of the
plane, the model is a chain of tetrahedra, linked on opposite edges.}
\label{novert}
\end{figure}

To start, consider two rows of atoms connected by horizontal and
diagonal bonds, as illustrated in Fig.~\ref{novert}.  The bonds
represent hopping terms, wherein an electron moves from one site to
another via a creation-annihilation operator pair in the Hamiltonian.
Later I will include vertical bonds, but for now consider just the
horizontal and diagonal connections.

Years ago during a course on quantum mechanics, I heard Feynman
present an amusing description of an electron's behavior when inserted
into a lattice.  If you place it initially on a single atom, the wave
function will gradually spread through the lattice, much like water
poured in a cell of a metal ice cube tray.  With damping, it settles
into the ground state which has equal amplitude on each atom.  To this
day I cannot fill an ice cube tray without thinking of this analogy
and pouring all the incoming water into a single cell.

I now complicate this picture with a magnetic field applied orthogonal
to the plane of the system.  This introduces phases as the electron
hops, causing interesting interference effects.  In particular,
consider a field of one-half flux unit per plaquette.  This means that
when a particle hops around a unit area (in terms of the basic lattice
spacing) the wave function picks up a minus sign.  Just where the
phases appear is a gauge dependent convention; only the total phase
around a closed loop is physical.  One choice for these phases is
indicated by the numbers on the bonds in Fig.~\ref{novert}.

\begin{figure}
\centerline{\includegraphics[width=.4\textwidth]{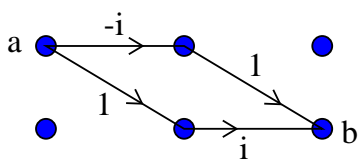}}
\caption{With half a unit of magnetic flux per plaquette, the paths
for an electron to move two sites interfere destructively.  
A particle on site $a$ cannot reach $b$.}
\label{cancel}
\end{figure}

The phase factors cause cancellations and slow diffusion.  For
example, consider the two shortest paths between the sites {\bf a} and
{\bf b} in Fig.~\ref{cancel}.  With the chosen flux, these paths
exactly cancel.  For the full molecule this cancellation extends to
all paths between these sites.  An electron placed on site {\bf a} can
never diffuse to site {\bf b}.  Unlike in the ice tray analogy, the
wave function will not spread to any site beyond the five nearest
neighbors.

\begin{figure}
\centerline{\includegraphics[width=.65\textwidth]{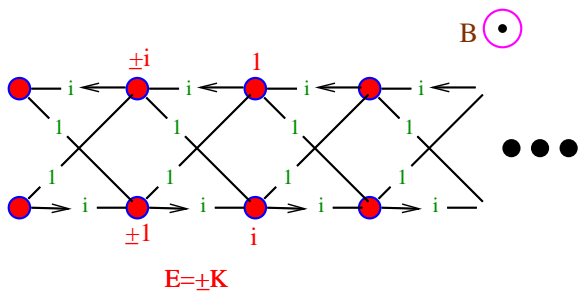}}
\caption{Two localized energy eigenstates occur on every plaquette of
the molecule.}
\label{novertone}
\end{figure}

As a consequence, the Hamiltonian has localized eigenstates.  While it
is perhaps a bit of a misuse of the term, these states are
``soliton-like'' in that they just sit there and do not change their
shape.  There are two such states per plaquette; one possible
representation for these two states is shown in Fig.~\ref{novertone}.
The states are restricted to the four sights labeled by their relative
wave functions.  Their energies are fixed by the size of the hopping
parameter $K$.

\begin{figure}
\centerline{\includegraphics[width=.65\textwidth]{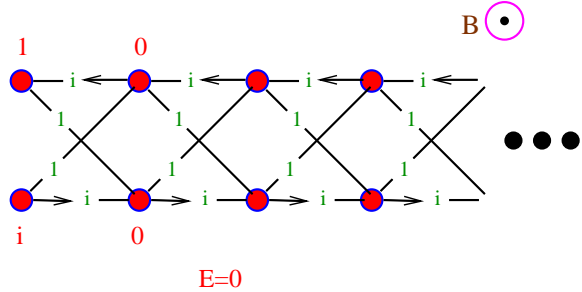}}
\caption{A zero energy state bound to the lattice end.}
\label{noverttwo}
\end{figure}

For a finite chain of length $L$ there are $2L$ atoms, and thus there
should be a total of $2L$ possible states for our electron (ignoring
spin).  There are $L-1$ plaquettes, and thus $2L-2$ of the above
soliton states.  This is almost the entire spectrum of the
Hamiltonian, but two states are left over.  These are zero energy
states bound to the ends of the system.  The wave function for one of
those is shown in Fig.~\ref{noverttwo}.  We now have the full spectrum
of the Hamiltonian: $L-1$ degenerate states of positive energy, a
similar number of degenerate negative energy states, and two states of
zero energy bound on the ends.

\begin{figure}
\centerline{\includegraphics[width=.65\textwidth]{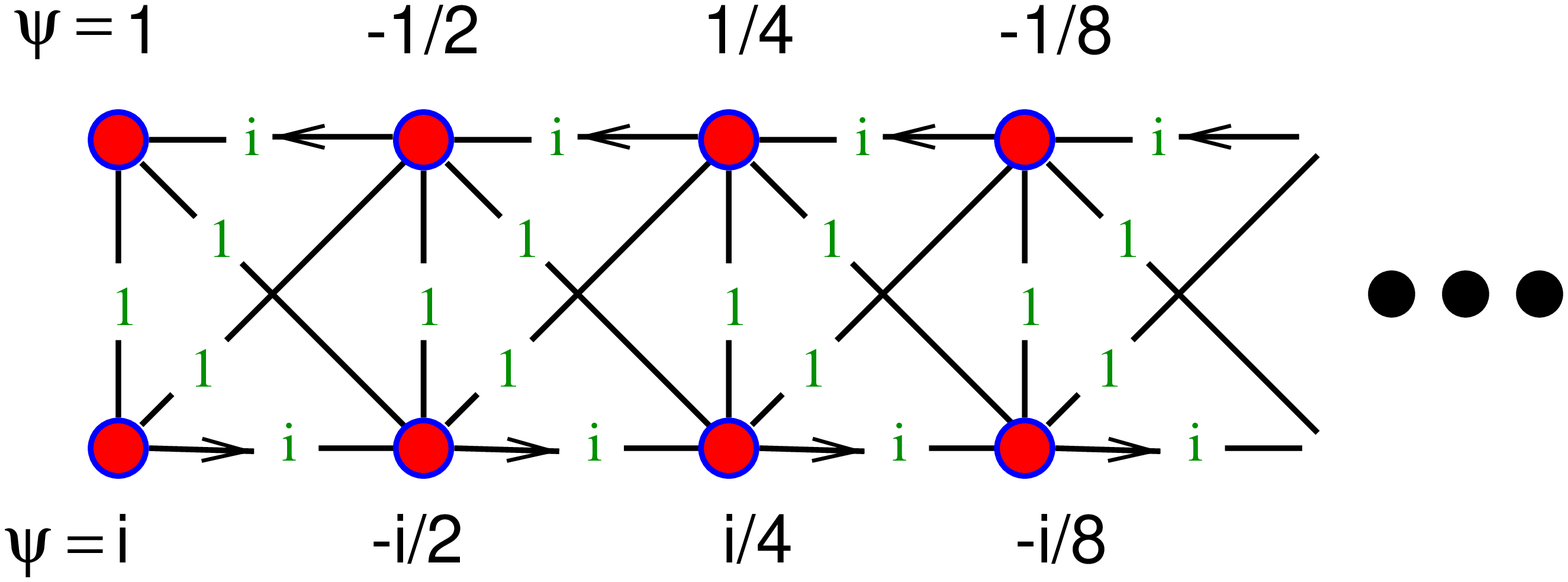}}
\caption{The zero energy state is robust under adding vertical bonds.}
\label{icetray3}
\end{figure}

Now consider what happens when vertical bonds are included in our
molecule.  The phase cancellations are no longer complete and the
solitonic states spread to form two bands, one with positive and one
with negative energy.  However, for our purposes, the remarkable
result is that the zero modes bound on the ends of the chain are
robust.  The corresponding wave functions are no longer exactly
located on the last atomic pair, but now have an exponentially
suppressed penetration into the chain.  Fig.~\ref{icetray3} shows the
wave function for one of these states when the vertical bond has the
same strength as the others.  There is a corresponding state on the
other end of the molecule.

When the chain is very long, both of the end states are forced to zero
energy by symmetry considerations.  First, since nothing distinguishes
one end of the chain from the other, they must have equal energy,
$E_L=E_R$.  On the other hand, a change in phase conventions,
effectively a gauge change, can change the sign of all the vertical
and diagonal bonds.  Following this with a left right flip of the
molecule will change the signs of the horizontal bonds.  This takes
the Hamiltonian to its negative, and shows that the states must have
opposite energies, $E_L=-E_R$.  This is indicative of a particle-hole
symmetry.  The combination of these results forces the end states to
zero energy, with no fine tuning of parameters.

For a finite chain, the exponentially decreasing penetration of the
end states into the molecule induces a small interaction between them.
They mix slightly to acquire exponentially small energies $E\sim \pm
e^{-\alpha L}$.  As the strength of the vertical bonds increases, so
does the penetration of the end states.  At a critical strength, the
mixing becomes sufficient that the zero modes blend into the positive
and negative energy bands.  In the full model, the mixing depends on
the physical momentum, and this disappearance of the zero modes is the
mechanism that removes the ``doublers'' when spatial momentum
components are near $\pi$ in lattice units \cite{mcih}.

\begin{figure}
\centerline{\includegraphics[width=.5\textwidth]{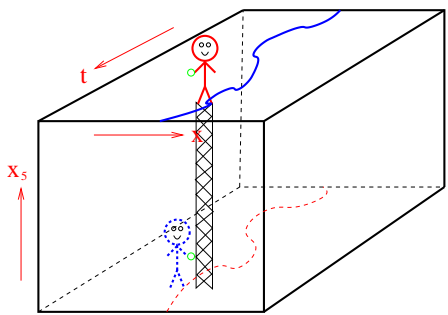}}
\caption{The zero modes of the chain molecules become the quarks of
which we are made.}
\label{kaplantwo}
\end{figure}

Energy levels forced to zero by symmetry lie at the core of the domain
wall fermion idea.  On every spatial site of a three dimensional
lattice we place one of these chain molecules.  The distance along the
chain is usually referred to as a fictitious ``fifth'' dimension.  The
different spatial sites are coupled, allowing particles in the zero
modes to move around.  These are the physical fermions.  The
symmetries that protect the zero modes now protect the masses of these
particles.  Their masses receive no additive renormalization, exactly
the consequence of chiral symmetry in the continuum.  The physical
picture is cartooned in Fig.~\ref{kaplantwo}, where I have rotated the
fifth dimension to the vertical.  Our world lines traverse the four
dimensional surface of this five dimensional manifold.

This scheme is for the fermions of the theory, and nothing extra is
needed for the gauge fields.  Indeed, we do not want the gauge fields
to see the extra dimension.  Thus we keep $A(x_\mu,x_5)=A(x_\mu)$
independent of $x_5$ and have no fifth component, i.e. $A_5=0$.  In
some sense calling our extra coordinate a dimension is a bit of a
convention; $x_5$ might as well be regarded as a ``flavor'' \cite{nn}.

The domain wall approach gives rise to a natural chiral theory on one
wall.  This gives a particularly elegant formulation of the strong
interactions, minimizing the doubling required by existing no-go
theorems.  In this picture the left and right handed quarks reside on
opposite walls.

For a chiral theory, however, the existence of anti-walls raises
unresolved questions.  For a finite fifth dimension the walls always
appear in pairs.  Because the gauge fields do not know about the fifth
dimension, the same gauge fields appear on each wall.  The opposite
chirality fermion zero modes found there represent ``mirror''
fermions; a theory with a left handed neutrino on one wall will
naturally have a right handed partner on the other.  How to resolve
this issue for the standard model is still controversial.

\begin{figure}
\centerline{\includegraphics[width=.6\textwidth]{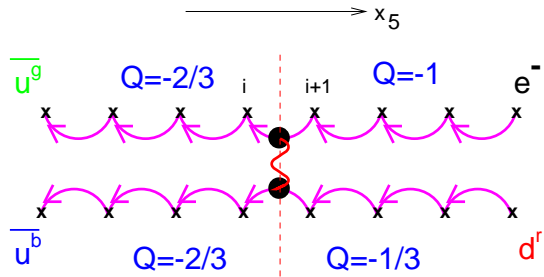}}
\caption
{Introducing a charge transfer involving four fermionic fields gives
rise to a possible scheme for putting the standard model on the
lattice.}
\label{transfer}
\end{figure}

One speculative approach was presented a few years ago \cite{smol},
where an unusual identification of the particles on the two walls was
enabled via the introduction of a four fermion coupling deep in the
interior of the extra dimension, as sketched in Fig.~\ref{transfer}.
The introduced four-fermion operator is ``technically irrelevant,''
and fully gauge invariant.  It is baryon number violating, but, as
noted earlier, this is a necessary feature of any fully finite
formulation of the standard model.

This particular approach has not received much attention because of
difficulties in treating the four fermion coupling.  In particular,
there is a serious danger that such a coupling could induce a
spontaneous breaking of one of the gauge symmetries.  This would be a
disaster for the picture since such breaking would naturally be at the
scale of the cutoff.

\begin{figure}
\centerline{\includegraphics[width=.3\textwidth]{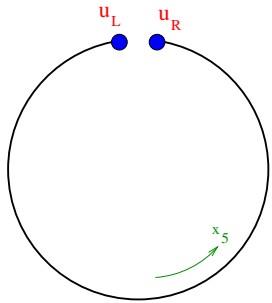}}
\caption {A compact fifth dimension must be cut in order to generate
the chiral zero modes of the domain-wall formalism.}
\label{circleone}
\end{figure}

I hope this description of domain-wall fermions in terms of simple
chain molecules has at least been thought provoking.  I now ramble on
with some general remarks about the basic scheme.  The existence of
the end states relies on using open boundary conditions in the fifth
direction.  If we were to curl our extra dimension into a circle, they
will be lost.  To retrieve them, consider cutting such a circle, as in
Fig.~\ref{circleone}.  Of course, if the size of the extra dimension is
finite, the modes mix slightly.  This is crucial for the scheme to
accommodate anomalies \cite{mcih}.

\begin{figure}
\centerline{\includegraphics[width=.3\textwidth]{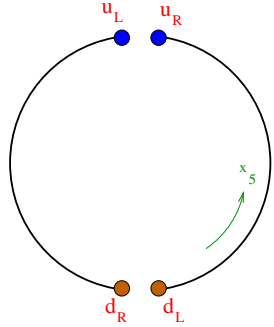}}
\caption {Cutting the compact fifth dimension twice
gives two flavors of fermion.  With the identifications here,
one flavored chiral symmetry is exact, even when the lattice
spacing and the size of the fifth dimension are finite.}
\label{circletwo}
\end{figure}

Suppose I want a theory with two flavors of light fermion, such as the
up and down quarks.  For this one might cut the circle twice, as shown
in Fig.~\ref{circletwo}.  Remarkably, this construction keeps one chiral
symmetry exact, even if the size of the fifth dimension is finite.
Since the cutting divides the molecule into two completely
disconnected pieces, in the notation of the figure we have the number
of $u_L+d_R$ particles absolutely conserved.  Similarly with
$u_R+d_L$.  Subtracting, we discover an exactly conserved axial charge
corresponding to the continuum current
$$
j_{\mu 5}^3 = \overline \psi \gamma_\mu\gamma_5 \tau^3 \psi
$$
The conservation holds even with finite $L_5$.  There is a small
flavor breaking since the $u_L$ mixes with the $d_R$.  These
symmetries are reminiscent of Kogut-Susskind \cite{kogutsusskind}, or
staggered, fermions, where a single exact chiral symmetry is
accompanied by a small flavor breaking.  Now, however, the extra
dimension gives additional control over the latter.

Despite this analogy, the situation is physically somewhat different
in the zero applied mass limit.  Staggered fermions are expected to
give rise to a single zero mass Goldstone pion, with the other pions
acquiring mass through the flavor breaking terms.  In my double cut
domain-wall picture, however, the zero mass limit has three degenerate
equal mass particles as the lowest states.  To see how this works it
is simplest to discuss the physics in a chiral Lagrangian language.
The finite fifth dimension generates an effective mass term, but it is
not in a flavor singlet direction.  It is in a flavor direction
orthogonal to the naive applied mass.  In the usual Mexican hat
picture, the two mass terms compete and the true vacuum rotates around
from the conventional ``sigma'' direction to the ``pi'' direction.

\begin{figure}
\centerline{\includegraphics[width=.4\textwidth]{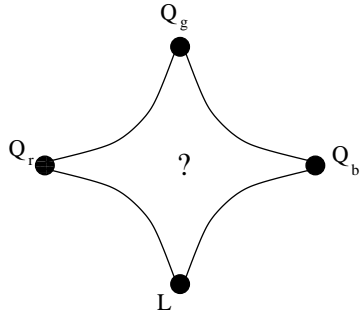}}
\label{starbw}
\caption {Perhaps all fermions are special modes of a single
higher-dimensional field.  Here the three quark fields might represent
different values of the internal $SU(3)$ symmetry, and $L$ could
represent a lepton from the same family.}
\end{figure}

Now I become more speculative.  The idea of using multiple cuts in the
fifth dimension to obtain several species suggests extensions to zero
modes on more complicated manifolds.  By having a variety of zero
modes, we have a mechanism to generate multiple flavors.  Maybe all
the physical fermions in four dimensions arise from a single fermion
field in the underlying higher dimensional theory.  Schematically we
might have something like shown in Fig.~\ref{starbw}.  where each
point represents some four dimensional surface and the question remark
represents structures in the higher dimension that need specification.

One nice feature provided by such a scheme is a possible mechanism for
the transfer of various quantum numbers involved in anomalous
processes.  For example, the baryon non-conserving 't Hooft
process\cite{thooft} might arise from a lepton flavor tunneling into
the higher manifold and reappearing on another surface as a baryon.
This generic mechanism is in fact the basis of the specific proposed
formulation of the standard model on the lattice\cite{smol} mentioned
earlier.

To summarize, I have argued that because it is totally finite, the
lattice forces honesty in understanding any peculiar phenomena that
arises, and this can reveal deep features of quantum field theory.
Chiral symmetry issues represent a dramatic example of this.

I presented a simple molecular picture for zero modes protected by
symmetry.  This illustrates the mechanism for mass protection in the
domain-wall formulation of lattice fermions.  Finally I speculated on
schemes for generating multiple fermion species from the geometry of
higher dimensional models.  The latter may have connections with the
activities in string theory.

\section*{Acknowledgment}
This manuscript has been authored under contract number
DE-AC02-98CH10886 with the U.S.~Department of Energy.  Accordingly,
the U.S. Government retains a non-exclusive, royalty-free license to
publish or reproduce the published form of this contribution, or allow
others to do so, for U.S.~Government purposes.

\end{document}